\documentclass[aps,pra,showpacs,superscriptaddress,preprint]{revtex4}
\usepackage{graphicx}

\catcode`ð=\active
 \defð{\u{g}}
 \catcode`Ð=\active
 \defÐ{\u{G}}
 \catcode`Ý=\active
 \defÝ{\. I}
 \catcode`ö=\active
 \defö{\"{o}}
 \catcode`Ö=\active
 \defÖ{\"O}
 \catcode`ü=\active
 \defü{\"{u}}
 \catcode`Ü=\active
 \defÜ{\"{U}}
 \catcode`Þ=\active
 \defÞ{\c{S}}
 \catcode`þ=\active
 \defþ{\c{s}}
 \catcode`ý=\active
 \defý{{\i}}
 \catcode`ç=\active
 \defç{\d{c}}
 \catcode`Ç=\active
 \defÇ{\d{C}}

\begin{document}

\title{Approximate $\ell$-State Solutions of a Spin-$0$ Particle
for Woods-Saxon Potential}

\author{\small Altuð Arda}
\email[E-mail: ]{arda@hacettepe.edu.tr}\affiliation{Department of
Physics Education, Hacettepe University, 06800, Ankara,Turkey}
\author{Ramazan Sever}
\email[E-mail: ]{sever@metu.edu.tr}\affiliation{Department of
Physics, Middle East Technical University, 06800, Ankara,Turkey}


\begin{abstract}
The radial part of Klein-Gordon equation is solved for the
Woods-Saxon potential within the framework of an approximation to
the centrifugal barrier. The bound states and the corresponding
normalized eigenfunctions of the Woods-Saxon potential are
computed by using the Nikiforov-Uvarov method. The results are
consistent with the ones obtained in the case of generalized
Woods-Saxon potential. The solutions of the Schr\"{o}dinger
equation by using the same approximation are also studied as a
special case, and obtained the consistent
results with the ones obtained before.
\end{abstract}

\maketitle

\newpage

\section{Introduction}
In recent years, the exact or approximate solutions of wave
equations have received a great attention. The Woods-Saxon (WS)
potential given by [1]

\begin{eqnarray}
V(r)=\,-\,\frac{V_0}{1+e^{(r-r_0)/a}}\,,
\end{eqnarray}
is one of the much studied potential in quantum mechanical
problems [2]. $V_0$ is the potential depth, the parameter $a$ is
the thickness of surface, and we denote the width of the potential
by $r_0$\,, which is proportional with target mass number $A$. The
coupled-channels approach is a powerful tool to reproduce in
heavy-ion physics, and Woods-Saxon potential, as a internuclear
potential, has an important role in the coupled-channels
calculations [2]. Further, the nuclear optical-model potential is
widely used to study elastic scattering processes of nucleons, and
heavy particles, and generate distorted waves in nuclear
reactions. The Woods-Saxon potential is one of the three parts of
nucleon-nucleus potential in the view of optical-model [3].

The wave equations with the Woods-Saxon potential can be solved
analytically for $s$-waves due to the centrifugal potential
barrier, and these solutions including the wave functions have
been obtained by using different methods [4, 9]. In this work, we
give the energy eigenvalues and the corresponding eigenfunctions
of the radial Klein-Gordon (KG) equation for usual Woods-Saxon
potential for any $\ell$ values by using an approximate term
instead of the centrifugal potential barrier. We solve the radial
part of KG equation by using Nikiforov-Uvarov (NU) method, which
is a powerful method to solve the second-order, linear
differential equations [10]. Further, we also study the energy
spectrum, and the corresponding wave functions of the
Schr\"{o}dinger equation for any $\ell$ value by using the same
approximation. So, we check out the consistency of our new
approximation scheme in the non-relativistic case.

The work is organized as follows. In Section II, we solve the
radial part of KG equation for usual Woods-Saxon potential for any
$\ell$ state by using an approximate potential term replaced by
centrifugal potential barrier. We find out the eigenvalues and
corresponding normalized eigenfunctions, and also give the results
for s-waves. We also apply our approximation to the case of the
Schr\"{o}dinger equation, and it makes possible to control the
accuracy of our approximation in non-relativistic region. We
summarize our concluding remarks in Section III.
\section{Bound States and Nikiforov-Uvarov Method}
In spherical coordinates, the radial part of KG equation can be
written as [11]

\begin{eqnarray}
\Big\{-\frac{\hbar^2}{2m_0}\,\frac{d^2}{dr^2}+\frac{\hbar^2\ell(\ell+1)}{2m_{0}r^2}+\frac{1}{2m_{0}c^2}
[m^2_{0}c^{4}-(E-V(r))^2]\Big\}\phi(r)=0\,,
\end{eqnarray}
where $E$ is the energy of the particle, $m_{0}$ is the rest mass
of particle, and $\ell$ is the angular momentum quantum number.

Let us write the potential as

\begin{eqnarray}
V(x)=\,-\,\frac{V_0}{1+e^{\beta x}}\,,
\end{eqnarray}
where $x=r-r_0$, and $\beta$ is a short notation, i.e.,
$\beta=1/a$. Eq. (2) can not be solved exactly because of the
centrifugal potential, but the term can be expand about $x=0$ as
the following

\begin{eqnarray}
V_{1}(r)=\,\frac{\hbar^2\ell(\ell+1)}{2m_{0}r^2}\,=\,\frac{D}{(1+\frac{x}{r_{0}})^2}=
D(1-2\frac{x}{r_{0}}+3(\frac{x}{r_{0}})^2+ \ldots)\,,
\end{eqnarray}
because the nuclear distance $r$ can not fluctuate very far from
the equilibrium for rather high vibrational levels [12, 13, 14],
which means that the series expansion in Eq. (4) is valid for
small $x$ values. The parameter $D$ in the above equation is a
short notation, i.e.,
$D=\,\frac{\hbar^2\ell(\ell+1)}{2m_{0}r^2_0}$\,.

We prefer the following form instead of the centrifugal potential
barrier

\begin{eqnarray}
V'_1(x)=\Bigg[D_0\,+\,\frac{D_1}{1+e^{\beta x}}\,
+\,\frac{D_2}{(1+e^{\beta x})^2}\,\Bigg]D\,,
\end{eqnarray}
where we use three new parameters $D_0, D_1$\,, and $D_2$\,. The
parameter $D_0$ corresponds to the constant term in series
expansion in Eq. (4), and the remaining two terms $D_1, D_2$
correspond to term proportional with $1/(1+e^{\beta x})$\,, and
proportional with $1/(1+e^{\beta x})^2$ in Eq. (2), respectively.

Expanding the potential $V'_1(x)$ around $x=0$ under the same
condition, and than combining equal powers with Eq. (4), one can
find the arbitrary constants $D_i (i=0, 1, 2)$ in the new form of
the potential as following

\begin{eqnarray}
D_0&=&\,\frac{12}{\beta^2 r^2_0}\,-\,\frac{4}{\beta
r_0}+1\,,\nonumber\\
D_1&=&\,-\,\frac{48}{\beta^2 r^2_0}\,+\,\frac{8}{\beta
r_0}\,,\nonumber\\ D_2&=&\,\frac{48}{\beta^2 r^2_0}\,.
\end{eqnarray}

So, we get two different 'effective' potentials of the form

\begin{eqnarray}
V_{eff}(x)&=&-\frac{V_{0}}{1+e^{\beta
x}}+\frac{\hbar^2\ell(\ell+1)}{2m_{0}r^2}\,,\\
V'_{eff}(x)&=&-\frac{V_{0}}{1+e^{\beta
x}}+DD_0+\frac{DD_1}{1+e^{\beta x}}+\frac{DD_2}{(1+e^{\beta
x})^2}\,.
\end{eqnarray}

In Figs. 1, and 2, we plot the variation of the 'effective'
potentials $V_{eff}(x)$\,, and $V'_{eff}(x)$ with respect to
$\beta x$ for three different $\ell$ values. The Figs. 1, and 2
show that there is a well consistency between of the potential
$V_{eff}(x)$\,, and our approximation in Eq. (5), where we use the
numerical values $V_{0}=43.1$ MeV, $a_{p}=0.67$ fm, $a_{a}=0.55$
fm, $r_{0}=3.44731$ fm, $m_{p}=1.007825$ amu, $m_{a}=1.00866$ amu
for particle, and antiparticle, respectively [19]. We see that the
deviation from the effective potential $V_{eff}(x)$ appears for
higher $\ell$ values ($\ell \geq 5$) for the range starting from
$\beta(r-r_{0}) > \sim2.5$\,. So, if we use the geometric average
value $A=56$ [20], we get an average as $r_{0}=4.91623$\,, and
taking $a=0.654$ fm [20], we could set an upper limit such as $r
<\,\sim 1.70974$ fm for a good consistency of our approximation.

Substituting Eq. (5) into Eq. (2), we get

\begin{eqnarray}
\Bigg\{\frac{d^2}{dx^2}&+&[\delta^2(E^2-m^2_{0}c^4)-\frac{2m_{0}}{\hbar^2}DD_{0}]\nonumber\\
&+&[2\delta^2EV_{0}-\frac{2m_{0}}{\hbar^2}DD_{1}]\frac{1}{1+e^{\beta
x}}+[\delta^2V^2_{0}-\frac{2m_{0}}{\hbar^2}DD_{2}]\frac{1}{(1+e^{\beta
x})^2}\Bigg\}\phi(x)=0\,,
\end{eqnarray}

By using the transformation $z=2(1+e^{\beta x})^{-1}$\,\,\,\,$(0
\leq x \leq \infty \rightarrow 0 \leq z \leq 1)$ , we have

\begin{eqnarray}
\frac{d^2\phi(z)}{dz^2}\,+\,\frac{2(1-z)}{z(2-z)}\,\frac{d\phi(z)}{dz}\,+\,\frac{1}{[z(2-z)]^2}
\left[-a_1^2z^2-a_2^2z-a_3^2\right]\phi(z)=0\,.
\end{eqnarray}
where

\begin{eqnarray}
-a^2_1&=&\,\frac{1}{\beta^2}\,(\delta^2V^2_{0}-2m_{0}DD_{2}/\hbar^2)\,,\nonumber\\
-a^2_2&=&\,\frac{2}{\beta^2}\,(2\delta^2EV_{0}-2m_{0}DD_{1}/\hbar^2)\,,\nonumber\\
-a^2_3&=&\,\frac{4}{\beta^2}\,(\delta^2(E^2-m^2_{0}c^4)-2m_{0}DD_{0}/\hbar^2)\,.
\end{eqnarray}

To apply the NU-method, we rewrite Eq. (10) in the following form

\begin{eqnarray}
\phi^{\prime\prime}(z)+\,\frac{\tilde{\tau}(z)}{\sigma(z)}\,\phi^{\prime}(z)
+\,\frac{\tilde{\sigma}(z)}{\sigma^2(z)}\,\phi(z)=0,
\end{eqnarray}
where $\sigma(z)$ and $\tilde{\sigma}(z)$ are polynomials with
second-degree, at most, and $\tilde{\tau}(z)$ is a polynomial with
first-degree. By using the following transformation for the total
wave function

\begin{eqnarray}
\phi(z)=\xi(z)\psi(z)
\end{eqnarray}
we get a hypergeometric type equation

\begin{eqnarray}
\sigma(z)\psi^{\prime\prime}(z)+\tau(z)\psi^{\prime}(z)+\lambda\psi(z)=0,
\end{eqnarray}
where $\xi(z)$ satisfies the equation

\begin{eqnarray}
\xi^{\prime}(z)/\xi(z)=\pi(z)/\sigma(z).
\end{eqnarray}
and the other part, $\psi(z)$, is the hypergeometric type function
whose polynomial solutions are given by

\begin{eqnarray}
\psi_n(z)=
\,\frac{b_n}{\rho(z)}\,\frac{d^n}{dz^n}[\sigma^n(z)\rho(z)],
\end{eqnarray}
where $b_n$ is a normalization constant, and the weight function
$\rho(z)$ must satisfy the condition

\begin{eqnarray}
\frac{d}{dz}[\sigma(z)\rho(z)]=\tau(z)\rho(z).
\end{eqnarray}
The function $\pi(z)$ and the parameter $\lambda$ required for
this method are defined as follows

\begin{eqnarray}
\pi(z)=\,\frac{\sigma^{\prime}(z)-\tilde{\tau}(z)}{2}\,
\pm\,\sqrt{(\frac{\sigma^{\prime}(z)-\tilde{\tau}(z)}{2})^2-\tilde{\sigma}(z)+k\sigma(z)}\,,
\end{eqnarray}
\begin{eqnarray}
\lambda=k+\pi^{\prime}(z)
\end{eqnarray}
The constant $k$ is determined by imposing a condition such that
the discriminant under the square root should be zero. Thus one
gets a new eigenvalue equation

\begin{eqnarray}
\lambda&=&\lambda_n=-n\tau^{\prime}-\,\frac{n(n-1)}{2}\,\sigma^{\prime\prime}\,,
(n=0, 1, 2, \ldots)
\end{eqnarray}
where

\begin{eqnarray}
\tau(z)=\tilde{\tau}(z)+2\pi(z)\,.
\end{eqnarray}
and the derivative of $\tau(z)$ must be negative.

Comparing Eq. (10) with Eq. (12), we have

\begin{eqnarray}
\tilde{\tau}(z)=2(1-z)\,,\,\,\,\,\,\sigma(z)=z(2-z)\,,\,\,\,\,\,
\tilde{\sigma}(z)=-a_1^2z^2-a_2^2z-a_3^2
\end{eqnarray}

Substituting this into Eq. (18), we get

\begin{eqnarray}
\pi(z)=\pm\sqrt{(a_1^2-k)z^2+(a_2^2+2k)z+a_3^2}.
\end{eqnarray}

The constant $k$ can be determined by the condition that the
discriminant of the expression under the square root has to be
zero

\begin{eqnarray}
(a_2^2+2k)^2-4a_3^2(a_1^2-k)=0\,.
\end{eqnarray}

The roots of $k$ are
$k_{1,2}=\,-\,\frac{1}{2}\,a_2^2\,-\,\frac{1}{2}\,a_3^2\mp\,\frac{1}{2}\,a_3A$,
where $A=\sqrt{a_3^2+2a_2^2+4a_1^2}$. Substituting these values
into Eq. (23), we get for $\pi(z)$ for $k_1$

\begin{eqnarray}
\pi(z)=\mp
\Big[\Big(\,\frac{A}{2}\,-\,\frac{a_3}{2}\,\Big)z+a_3\Big]\,,
\end{eqnarray}
and for $k_2$

\begin{eqnarray}
\pi(z)=\mp
\Big[\Big(\,\frac{A}{2}\,+\,\frac{a_3}{2}\,\Big)z-a_3\Big]\,,
\end{eqnarray}

Now we find the polynomial $\tau(z)$ from $\pi(z)$ for $k_2$

\begin{eqnarray}
\tau(z)=2+2a_3-2\Big(\,\frac{A}{2}\,+\,\frac{a_3}{2}\,+1\,\Big)z.
\end{eqnarray}
so its derivative
$-2\Big(\,\frac{A}{2}\,+\,\frac{a_3}{2}\,+1\,\Big)$ is negative.
We have from Eq. (19)

\begin{eqnarray}
\lambda=\,-\,\frac{1}{2}\,\Big(a_2^2+a^2_3+Aa_3+A+a_3\Big)\,,
\end{eqnarray}
and Eq. (20) gives us

\begin{eqnarray}
\lambda_n=2n\Big(\,\frac{A}{2}\,+\,\frac{a_3}{2}\,+1\,\Big)+n^2-n\,.
\end{eqnarray}
Substituting the values of the parameters given by Eq. (11), and
setting $\lambda=\lambda_n$, one can find the energy eigenvalues
for any $\ell$-state

\begin{eqnarray}
E_{n,\ell}&=&-\frac{[\beta^2n^2_1+4\delta^2V^2_0-\frac{8m_{0}D}{\hbar^2}(D_{1}+D_{2})]V_{0}}
{2(\beta^2n^2_{1}+4\delta^2V^2_{0})}\nonumber\\
&\pm&\frac{\beta
n_{1}}{\delta}\sqrt{\,\frac{2m^2_{0}c^4\delta^2+\frac{2m_{0}D}{\hbar^2}
(2D_{0}+D_{1}+D_{2})}{2(\beta^2n^2_{1}+4\delta^2V^2_{0})}\,-\Bigg[\,\frac{\frac{2m_{0}D}{\hbar^2}(D_{1}+D_{2})}
{\beta^2n^2_{1}+4\delta^2V^2_{0}}\,\Bigg]^2\,-\,\frac{1}{16}\,}\,,
\end{eqnarray}
where $n$ is the principal quantum number, and

\begin{eqnarray}
n_{1}=-(2n+1)+\sqrt{1+4a^2_1\,}\,.
\end{eqnarray}
From this result, we can easily get the energy spectra for
$s$-waves by setting $D=D_{0}=D_{1}=D_{2}=0$

\begin{eqnarray}
E_{n,\ell=0}=-\frac{V_{0}}{2}\pm\,\frac{\beta
n'_{1}}{\delta}\sqrt{\,\frac{m^2_{0}c^4\delta^2}{\beta^2n'^2_{1}+4\delta^2V^2_{0}}-\frac{1}{16}\,}\,,
\end{eqnarray}
where

\begin{eqnarray}
n'_{1}=-(2n+1)+\sqrt{1+4a'^2_1\,}\,.
\end{eqnarray}
where $a'_1=a_1(\ell=0)$.

It can be seen that the eigenvalues are real under the condition
that $\frac{2m^2_{0}c^4\delta^2+\frac{2m_{0}D}{\hbar^2}
(2D_{0}+D_{1}+D_{2})}{2(\beta^2n^2_{1}+4\delta^2V^2_{0})}>\Big[\,\frac{\frac{2m_{0}D}{\hbar^2}(D_{1}+D_{2})}
{\beta^2n^2_{1}+4\delta^2V^2_{0}}\,\Big]^2+\frac{1}{16}$\,.
Further, the parameters $D_{0}, D_{1}, D_{2}$ used to describe the
approximate potential form in Eq. (5) are real in the case.

We give the variation of the bound states energy of particle + a
nucleon with $A$, and antiparticle + a nucleon with $A$ with
respect to $n$ for different $\ell$ values for in Figs. 3, and 4,
respectively. We choose the target mass number as $A=20$\,, and
use the numerical values given above.

In order to find the eigenfunctions, we first compute the weight
function from Eq. (17)

\begin{eqnarray}
\rho(z)=z^{a_3}(2-z)^{A}\,,
\end{eqnarray}
and the wave function becomes

\begin{eqnarray}
\psi_{n\ell}\,(z)&=&\,\frac{b_n}{z^{a_3}(2-z)^{A}}\,\frac{d^n}{dz^n}\,\left[
\,z^{n+a_3}\,(2-z)^{n+A}\right]\,.
\end{eqnarray}
where $b_n$ is a normalization constant. The polynomial solutions
can be written in terms of the Jacobi polynomials [15-17]

\begin{eqnarray}
\psi_{n\ell}\,(z)=b_n\,P_n^{(a_3,\,\,
A)}\,(1-z)\,,\,\,\,\,\,A>-1\,,\,\,\,\,\,a_3>-1\,.
\end{eqnarray}
On the other hand, the other part of the wave function is obtained
from Eq. (15) as

\begin{eqnarray}
\xi(z)=z^{a_3/2}\,(2-z)^{A/2}\,.
\end{eqnarray}
Thus, the total eigenfunctions take

\begin{eqnarray}
\phi_{n\ell}\,(z)=b'_n\,(2-z)^{A/2}z^{a_3/2}\,P_n^{(a_3,\,\,A)}\,(1-z)\,.
\end{eqnarray}
where $b'_n$ is the new normalization constant. It is obtained
from

\begin{eqnarray}
\frac{2}{\beta}\int_{0}^{1}\Big(\frac{\left|\phi_{n\ell}(z)\right|}{\sqrt{z^{-1}(2-z)\,}}\Big)^2dz=1\,.
\end{eqnarray}
To evaluate the integral, we use the following representation of
the Jacobi polynomials [16, 17]

\begin{eqnarray}
P_n^{(\sigma,\,\varsigma)}(z)=\,\frac{\Gamma(n+\sigma+1)}{n!\Gamma(n+\sigma+\varsigma+1)}
\,\sum_{m=0}^{n}\,\frac{\Gamma(n+1)}{\Gamma(m+1)\Gamma(n-m+1)}\,
\frac{\Gamma(n+\sigma+\varsigma+m+1)}{\Gamma(m+\sigma+1)}\,(\,\frac{z-1}{2})^m\,,\nonumber\\
\end{eqnarray}
Hence, from Eq. (39), and with the help of Eq. (40), we get

\begin{eqnarray}
\frac{1}{\beta}[g(n,m)\times g(r,s)]
\left|b'_n\right|^2\int_{0}^{1}z^{m+s+a_3+1}\,(2-z)^{A-1}\,dz=1\,,
\end{eqnarray}
where $g(n,m)$, and $g(r,s)$ are two arbitrary functions of the
parameters $A$, and $a_3$, and given by

\begin{eqnarray}
g(n,m)&=&(-1)^{m+1/2}(-2)^{-m+1/2}\,\frac{\Gamma(a_3+n+1)}{n!\Gamma(A+a_3+n+1)}\nonumber\\&\times&
\,\sum_{m=0}^{n}\,\frac{\Gamma(n+1)}{\Gamma(m+1)\Gamma(n-m+1)}\,
\frac{\Gamma(A+a_3+n+m+1)}{\Gamma(a_3+A+1)}\,,\nonumber\\
\end{eqnarray}
and

\begin{eqnarray}
g(r,s)=g(n,m)\,(n \rightarrow r; m \rightarrow s)\,.
\end{eqnarray}
The integral in Eq. (41) can be evaluated by using the following
integral representation of hypergeometric type function
$_2F_1(a,b;c;z)$ [13]

\begin{eqnarray}
_2F_1(a,b;c;z)=\,\frac{\Gamma(c)}{\Gamma(b)\Gamma(c-b)}\,\int_{0}^{1}
t^{b-1}\,(1-t)^{c-b-1}\,(1-tz)^{-a}\,dt\,,
\end{eqnarray}
by setting the variable $z \rightarrow\,\frac{z}{2}$, and taking
$c=1+b$\,,\,$z=1$, one gets

\begin{eqnarray}
\int_{0}^{1}\,t^{b-1}\,(2-t)^{-a}\,dt=\,\frac{\Gamma(b)\Gamma(1)}{2\Gamma(1+b)}\,
_2F_1\,(a,b;1+b;\,\frac{1}{2}\,)\,,
\end{eqnarray}
From last equation

\begin{eqnarray}
\int_{0}^{1}\,z^{m+s+a_3+1}\,(2-z)^{A-1}\,dz&=&\,\frac{\Gamma(m+s+a_3+2)\Gamma(1)}{2\Gamma(m+s+a_3+3)}\nonumber\\
&\times&_2F_1(1-A,m+s+a_3+2;m+s+a_3+3;\,\frac{1}{2}\,)\,,\nonumber\\
\end{eqnarray}
where we set $b-1=m+s+a_3+1$\,, and $a=1-A$.

By using the following identities of hypergeometric type functions
[18]

\begin{eqnarray}
_2F_1\,(a,b;c;-1)&=&\,\frac{\Gamma\Big(\,\frac{1}{2}b+1\Big)\Gamma(b-a+1)}{\Gamma(b+1)
\Gamma(\Big(\,\frac{1}{2}b-a+1\Big)}\,,\,\,\,(a-b+c=1\,,\,\,b>0)\\
_2F_1\,(a,b;c;\frac{1}{2})&=&2^a\,_2F_1\,(a,c-b;c;-1)\,,
\end{eqnarray}
the hypergeometric type function of $_2F_1\,(a,b;c;z)$ in Eq. (46)
can be evaluated as

\begin{eqnarray}
_2F_1\,(1-A,b;1+b;\frac{1}{2})=\,\frac{\sqrt{\pi}}{2^{A}}\,\frac{\Gamma(1+A)}{\Gamma\Big(\,\frac{1}{2}\,+A\Big)}\,,
\,\,\,(A=b-1)\,,
\end{eqnarray}
Finally, we get the normalization constant as

\begin{eqnarray}
\left|b'_n\right|^2=\,\frac{2^{1+A}}{\beta\sqrt{\pi}}\,\frac{\Gamma(m+s+a_3+3)\Gamma\Big(\,\frac{1}{2}\,+A\Big)}
{\Gamma(m+s+a_3+2)\Gamma(1+A)[g(n,m) \times g(r,s)]}\,.
\end{eqnarray}

It is interested to study the approximation scheme of potential
given by Eq. (5) in the case of Schr\"{o}dinger equation (SE). It
also makes to possible to check out the results obtained by using
the approximation Eq. (5) in the non-relativistic region.

The radial part of SE is given by

\begin{eqnarray}
-\frac{\hbar^2}{2m_{0}}\frac{d^2\phi(r)}{dr^2}+\Bigg\{\frac{\hbar^2\ell(\ell+1)}{2m_{0}r^2}
+V(r)-E\Bigg\}\,\phi(r)=0\,,
\end{eqnarray}

By using the approximate expression of the centrifugal potential
barrier, and the transformation $z=(1+e^{\beta x})^{-1}$\,, we
have

\begin{eqnarray}
\frac{d^2\phi(z)}{dz^2}+\frac{2z-1}{z(z-1)}\,\frac{d\phi(z)}{dz}+\frac{1}{[z(z-1)]^2}
\left[-\epsilon^2-\gamma^2z-\kappa^2 z^2\right]\phi(z)=0\,.
\end{eqnarray}
where

\begin{eqnarray}
\epsilon^2&=&\frac{2m_{0}}{\beta^2\hbar^2}(DD_{0}-E)\,,\nonumber\\
\gamma^2&=&\frac{2m_{0}}{\beta^2\hbar^2}(DD_{1}-V_{0})\,,\nonumber\\
\kappa^2&=&\frac{2m_{0}}{\beta^2\hbar^2}DD_{2}\,.
\end{eqnarray}

Following the same procedure, the energy eigenvalues are written

\begin{eqnarray}
E_{n\ell}&=&\frac{\hbar^2\ell(\ell+1)}{2m_{0}r^2_{0}}D_{0}\nonumber\\&-&\frac{\hbar^2}{2m_{0}a^2}
\Bigg[\frac{1}{4}\Big[2n+1+\sqrt{1+\frac{4\ell(\ell+1)a^2}{r^2_{0}}D_{2}\,}\,\Big]
-\frac{\frac{\ell(\ell+1)a^2}{r^2_{0}}(D_{1}+D_{2})-\frac{2m_{0}V_{0}a^2}{\hbar^2}}{2n+1+
\sqrt{1+\frac{4\ell(\ell+1)a^2}{r^2_{0}}D_{2}\,}}\Bigg]^2\,.
\end{eqnarray}

We plot the variation of the energy eigenvalues obtained from Eq.
(54) with respect to $n$ for $\ell=0$ in Fig. 5. We give the
dependence of the energy spectrum to $n$ for the values $\ell=1,
2$ in Fig. 6 by using the numerical values given above.

We get easily the energy spectrum for $s$-waves

\begin{eqnarray}
E_{n,\ell=0}=-\frac{\hbar^2}{2m_{0}a^2}\Bigg[\Big(\frac{n+1}{2}\Big)^2+\Big(\frac{m_{0}V_{0}a^2}
{\hbar^2(n+1)}\Big)^2+\frac{m_{0}V_{0}a^2}{\hbar^2}\Bigg]\,.
\end{eqnarray}
which is consistent with obtained in Ref. [9].

The corresponding eigenfunctions are written

\begin{eqnarray}
\phi_{n\ell}(z)=a_n\,(1-z)^{-A}\,z^{-\epsilon}\,P_n^{(-2\epsilon\,,\,-2A\,)}\,(1-2z)\,.
\end{eqnarray}
where $A=\sqrt{\epsilon^2+\gamma^2+\kappa^2\,}$\,, and $a_n$ is
normalization constant. Let us discuss the behavior of the
approximate wave function at the origin. Taking into account that
$z=(1+e^{\beta x})^{-1}$ and $x=r-r_{0}$\,, we get the following
limits

\begin{eqnarray}
z=\left\{
\begin{array}{rl}
0 & \text{if}\,\,x \rightarrow \infty\\
\frac{1}{2} & \text{if}\,\,x \rightarrow 0\\
\end{array}\right.
\end{eqnarray}
We should take the limit $z \rightarrow \frac{1}{2}$ to discuss the
behaviour of the wave function at origin. We obtain the wave
function in this limit as

\begin{eqnarray}
\phi_{n\ell}\,(z)\sim(\frac{1}{2})^{-A-\epsilon}
P_n^{(-2\epsilon\,,\,-2A\,)}\,(1-2z \rightarrow 0)\,,
\end{eqnarray}
The Jocabi polynomials has the following form in this limit
[17,18]

\begin{eqnarray}
P_n^{(-2\epsilon\,,\,-2A\,)}\,(y)&=&\frac{1}{2^n}\sum_{k=0}^{n}\left(
\begin{array}{cc}
n-2\epsilon\\
k
\end{array}\right)
\left(\begin{array}{cc}
n-2A\\
n-k
\end{array}\right)(1+y)^{k}(y-1)^{n-k}\,,
\end{eqnarray}
where $y$ identifies $1-2z$\,. So we write the approximate wave
function in the first order of $z$

\begin{eqnarray}
&&\phi_{n\ell}\,(z)\nonumber\\&\sim&
\Big\{(-1)^{n-k}+\Big[(-1)^{n-k-1}(n-k)+(-1)^{n-k}k\Big]z+\ldots\Big\}
f(n,k,\epsilon,A)\,,
\end{eqnarray}
where

\begin{eqnarray}
f(n,k,\epsilon,A)=\frac{1}{2^n}\sum_{k=0}^{n}\left(
\begin{array}{cc}
n-2\epsilon\\
k
\end{array}\right)
\left(\begin{array}{cc}
n-2A\\
n-k
\end{array}\right)\,.
\end{eqnarray}

It is seen from Eq. (60) that the approximate wave function is
finite at the origin. It will have the same behavior with the
approximate solution.

As a final test, we compare our results obtained from Eq. (54)
with the ones given in Ref. [21], where the problem is solved
numerically, and the results are obtained by using a MATLAB
package called as the MATSLISE. So, we use the following form of
the Woods-Saxon potential in Eq. (3)

\begin{eqnarray}
V(x)=-\frac{50}{1+e^{(x-7)/0.6}}\,,
\end{eqnarray}
where we should stress that the variable $x$ in last equation
denotes the radial coordinate $r$ in the rest of the present work.

We obtain energy eigenvalues of the Woods-Saxon potential given in
Eq. (62) by using the approach in Eq. (5) numerically, and
summarize the results in Table I. We give the eigenfunctions
obtained without, and with using the approximation given in Eq.
(5) for the Woods-Saxon potential given in Eq. (62) in Fig. 7 for
$n=1$\,, and $\ell=1$\,, and Fig. 8 for $n=6$\,, and $\ell=5$\,,
respectively. The deviations between the results obtained with and
without the approximation increase as the values of $n$\,, and
$\ell$\ increase, and there is a shifting to the left between the
approximate and exact eigenfunctions. These are expected results,
because the validity of the approximation is specified in the
Figs. 1, and 2 for the range as ($r < \sim 1.70794$ fm), while we
set the molecular distance $r$ as $r_{0}=7$ in Eq. (62).

\section{Conclusion} We have solved analytically the radial part of the KG equation for
the usual Woods-Saxon potential in the framework of an
approximation to the centrifugal potential term for any $\ell$
values. The energy spectra and the corresponding wave functions
are obtained by applying the NU-method. We give in Figs. 1, and 2
the variations of $V_{eff}(r)$\,, and $V'_{eff}(r)$ with respect
to $\beta(r-r_0)$ for $\ell=1, 2 ,5$. We have pointed out that the
consistency between $V_{eff}$\,, and our new approximation is very
well, and reliable results can be obtained by using the
approximation scheme used in the present work. To check our
results, we have also calculated analytically the energy
eigenvalues of the particle and antiparticles for the $s$-waves.
We have found that the analytical results are consistent with
those in Ref. [4] if $q \rightarrow 1$. Further, we have also
studied independently the energy spectrum of the Schr\"{o}dinger
equation by using the same approximation to check out our results
in the non-relativistic region. We have seen that the results
obtained analytically for the case of the Schr\"{o}dinger equation
are the same for $s$-waves with Ref. [9]. Further, we give some
numerical results for the eigenvalues of the Schr\"{o}dinger
equation with different values of the quantum numbers $n$\,, and
$\ell$ in Table I. We compare the eigenfunctions for two different
$\ell$ values obtained with, and without using the new
approximation in Figs. 7-8.

\section{Acknowledgments}
This research was partially supported by the Scientific and
Technical Research Council of Turkey. The authors would like to
thank the referee for helpful comments, and also to F.~Schoeberl
to communicate with us about the numerical calculations by using a
MATHEMATICA package. The author A.~Arda would like to thank to
E.~Sorgun for her kind patience.

\newpage

\begin{table}
\caption{The energy eigenvalues of Woods-Saxon potential in Eq.
(57) derived in MATSLISE.}
\begin{ruledtabular}
\begin{tabular}{ccc}
 & Our Work & Numerical Solution \\
\hline
1s & -49.57 & -49.57 \\
2s & -48.50 & -48.50 \\
2p & -49.52 & -49.17 \\
3s & -46.96 & -46.96 \\
3p & -48.45 & -47.84 \\
3d & -49.40 & -48.68 \\
4s & -45.02 & -45.02 \\
4p & -46.91 & -46.09 \\
4d & -48.33 & -47.11 \\
4f & -49.22 & -48.12 \\
5s & -40.11 & -40.11 \\
5p & -44.96 & -43.96 \\
5d & -46.79 & -45.15 \\
5f & -48.16 & -46.32 \\
5g & -48.99 & -47.49 \\
6s & -37.21 & -37.21 \\
6p & -42.67 & -41.50 \\
6d & -44.85 & -42.85 \\
6f & -46.62 & -44.17 \\
6g & -47.93 & -45.48 \\
6h & -48.70 & -46.79 \\
\end{tabular}
\end{ruledtabular}
\end{table}

\newpage

\begin{figure}[htbp]
\centering
\includegraphics[height=3in, width=5in, angle=0]{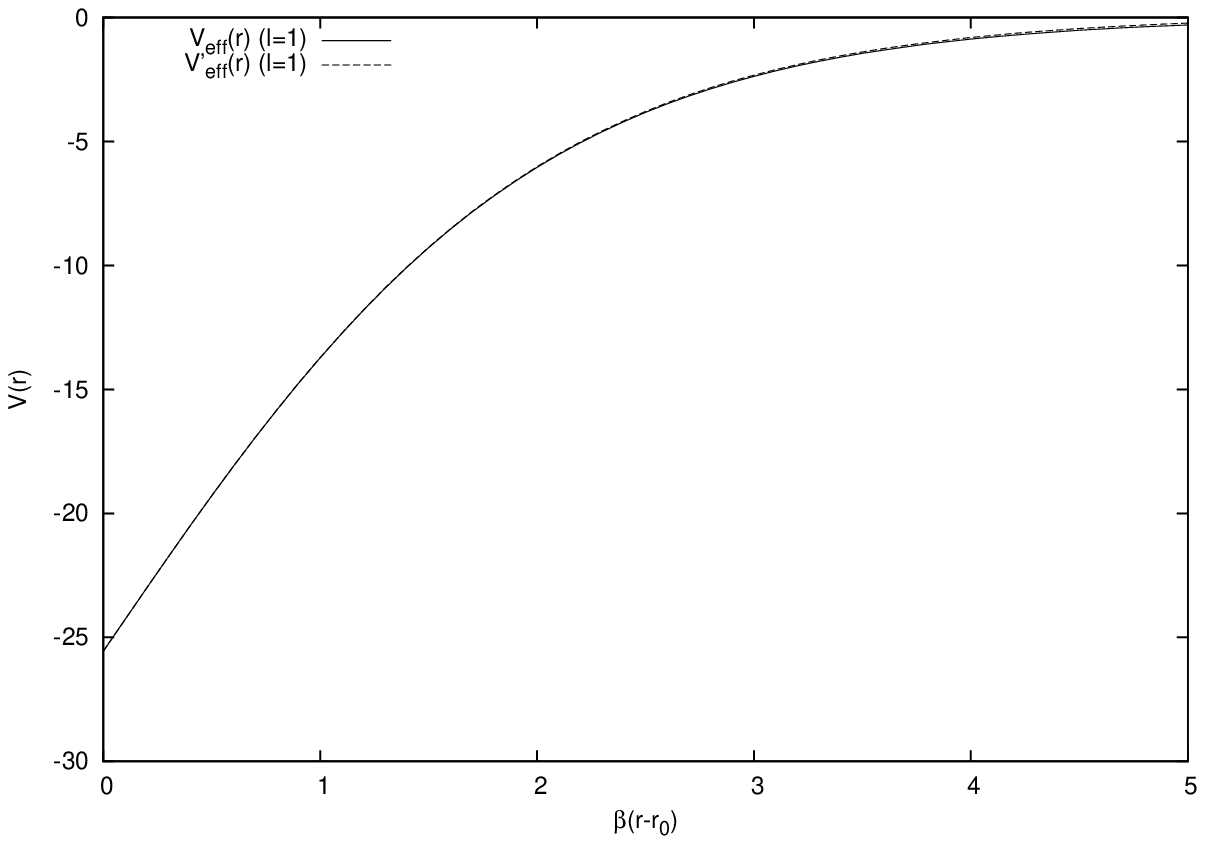}
\caption{The variations of $V_{eff}(r)$\,, and $V'_{eff}(r)$ with
respect to $\beta(r-r_0)$ for $\ell=1$.}
\end{figure}

\begin{figure}[htbp]
\centering
\includegraphics[height=3in, width=5in, angle=0]{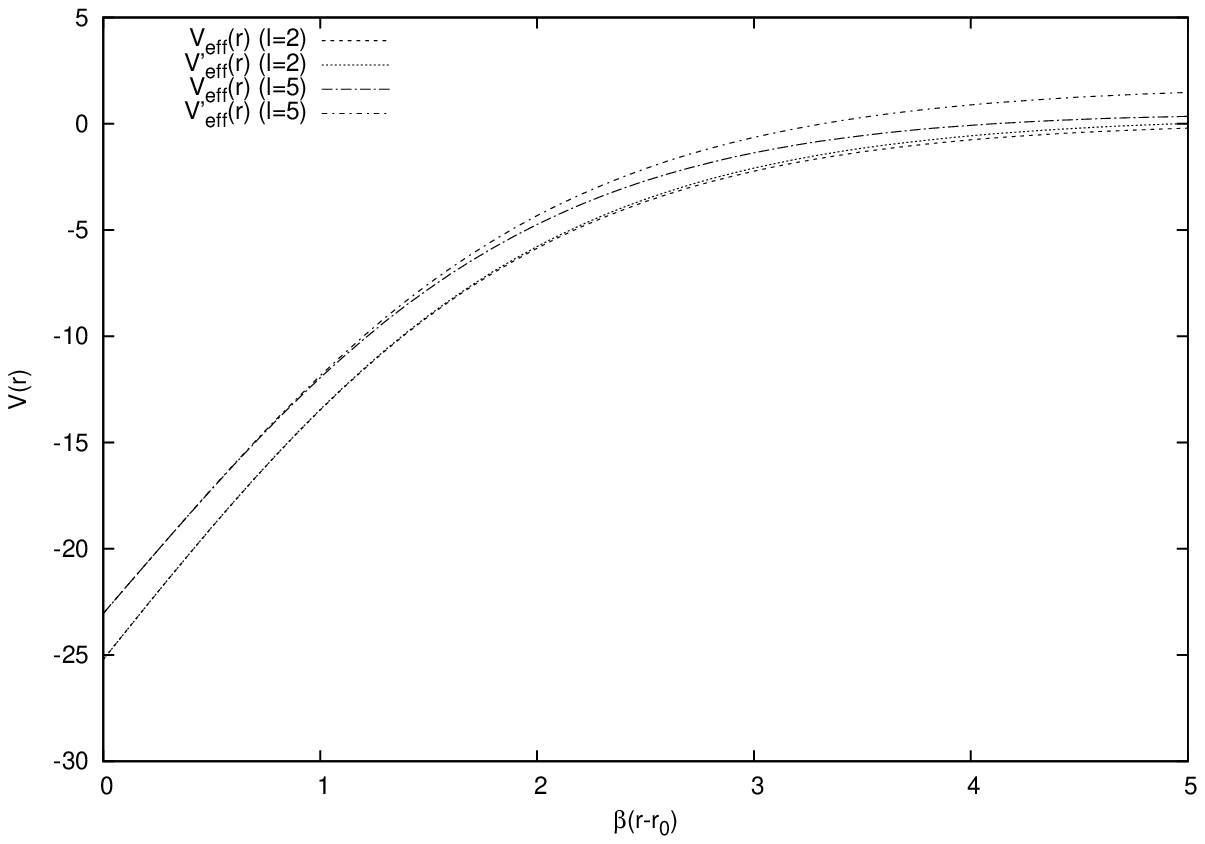}
\caption{The variations of $V_{eff}(r)$\,, and $V'_{eff}(r)$ with
respet to $\beta(r-r_0)$ for $\ell=2, 5$.}
\end{figure}

\newpage

\begin{figure}[htbp]
\centering
\includegraphics[height=2.5in, width=5in, angle=0]{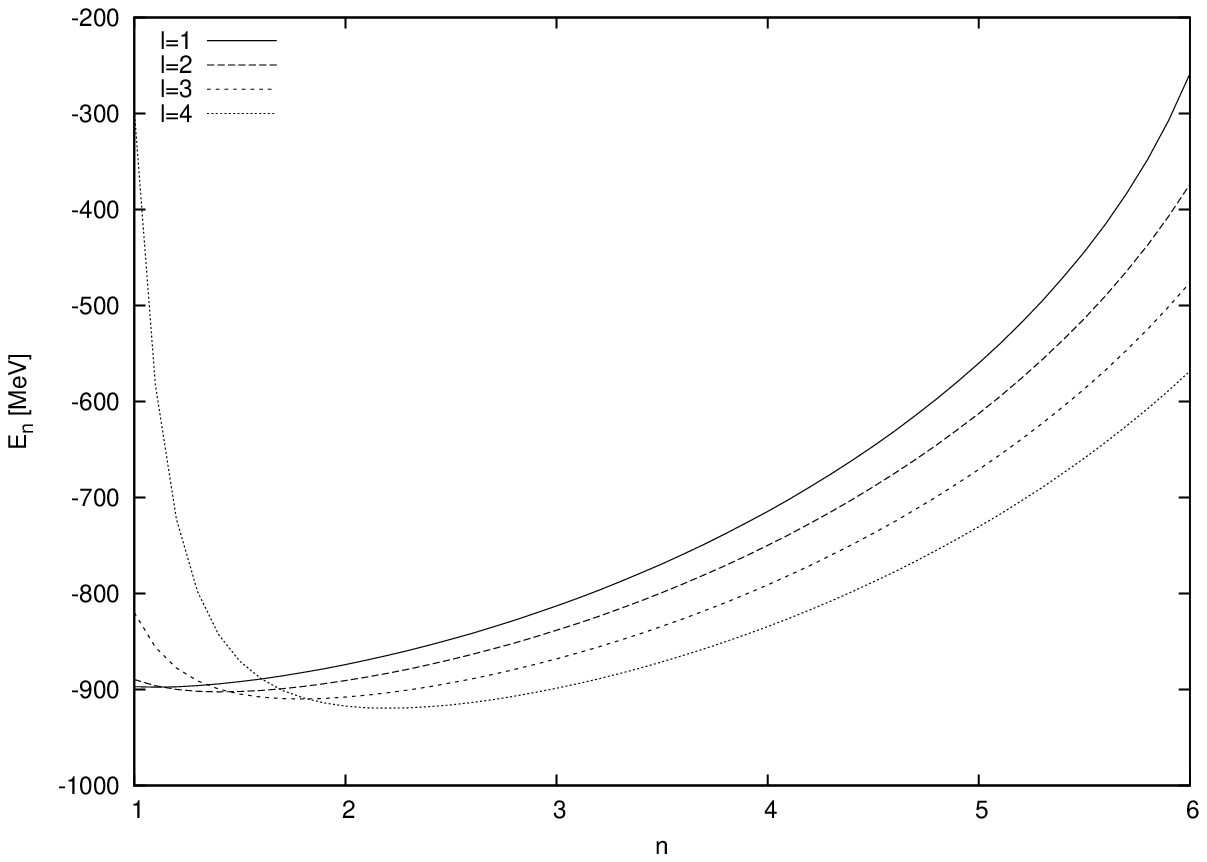}
\caption{The dependence of a particle energy levels in the case of
the Klein-Gordon to $n$ for different $\ell$ values $1, 2, 3, 4$.}
\end{figure}

\begin{figure}[htbp]
\centering
\includegraphics[height=2.5in, width=5in, angle=0]{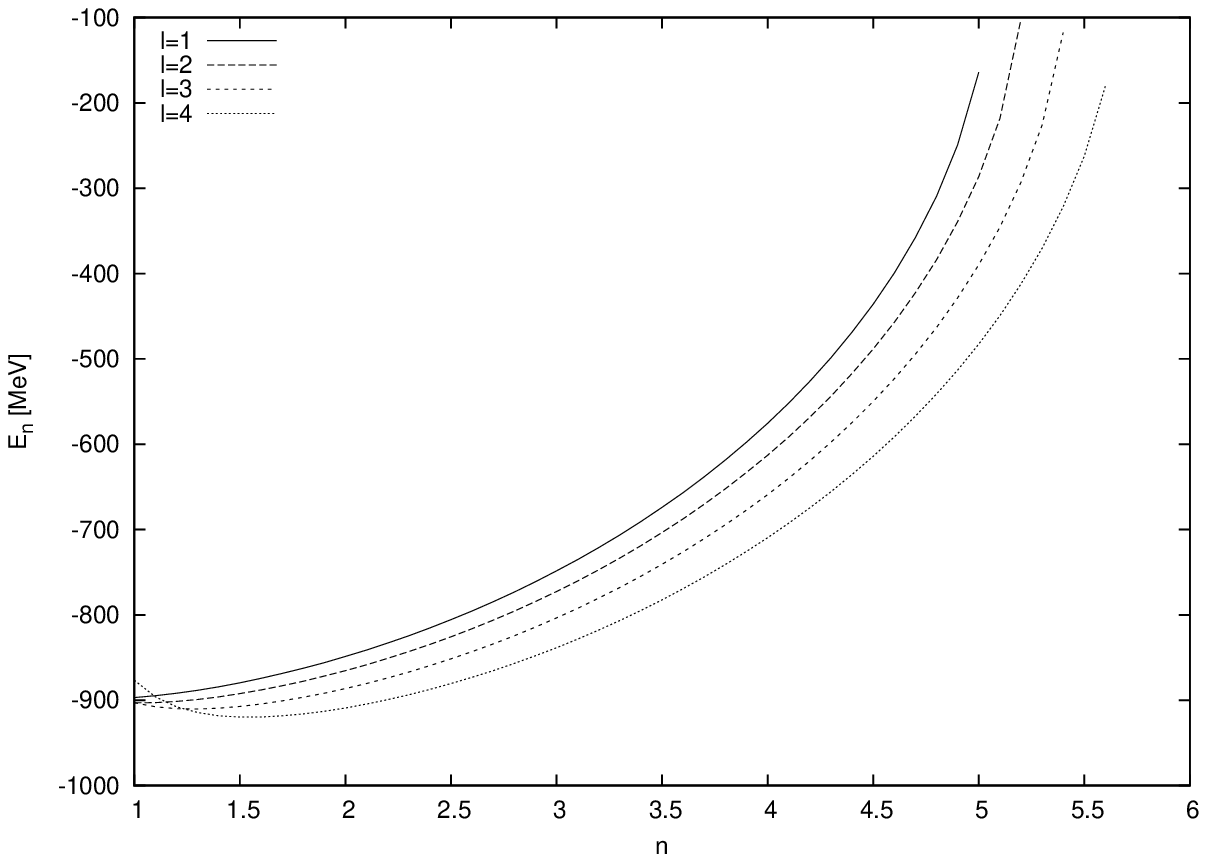}
\caption{The dependence of a antiparticle energy levels in the
case of the Klein-Gordon to $n$ for different $\ell$ values $1, 2,
3, 4$.}
\end{figure}

\newpage

\begin{figure}[htbp]
\centering
\includegraphics[height=2.5in, width=5in, angle=0]{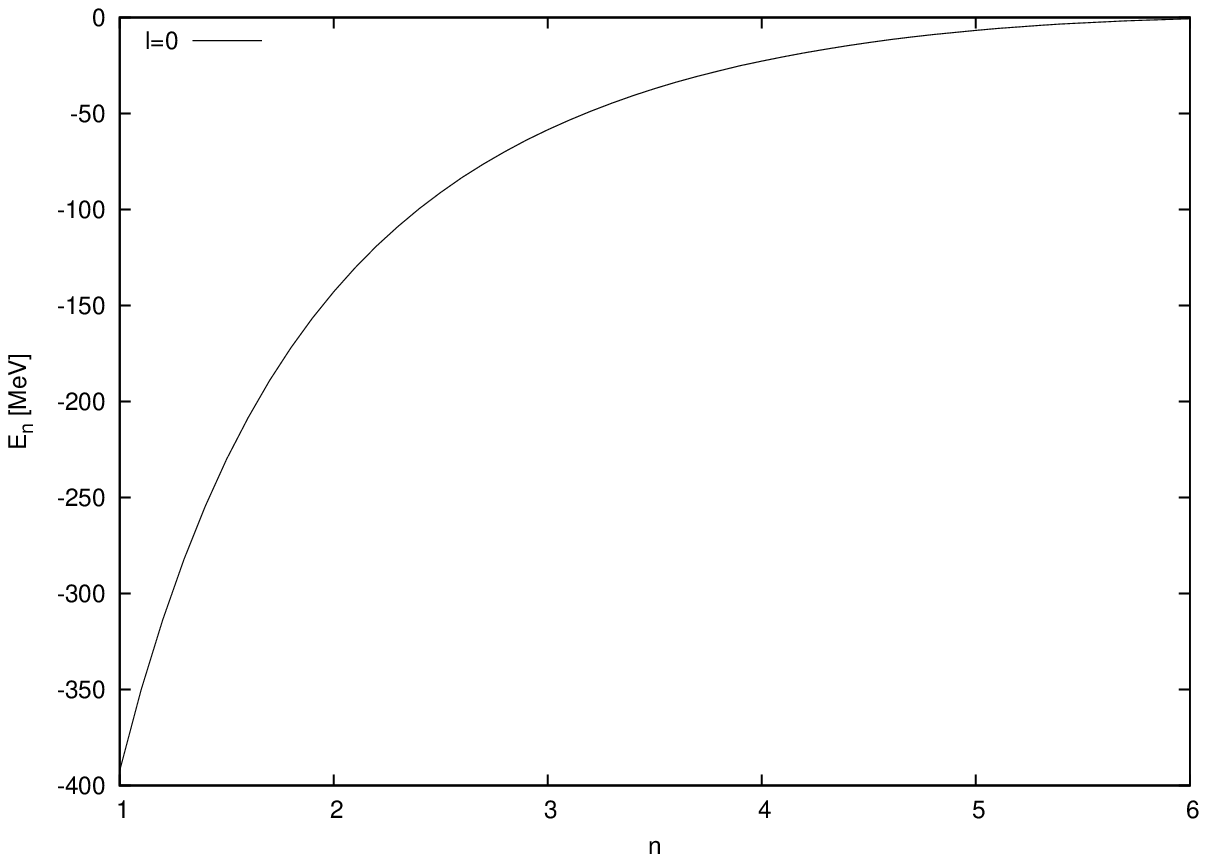}
\caption{The variation of the Schr\"{o}dinger energy level with
respect to $n$ for $\ell=0$\,.}
\end{figure}

\begin{figure}[htbp]
\centering
\includegraphics[height=2.5in, width=5in, angle=0]{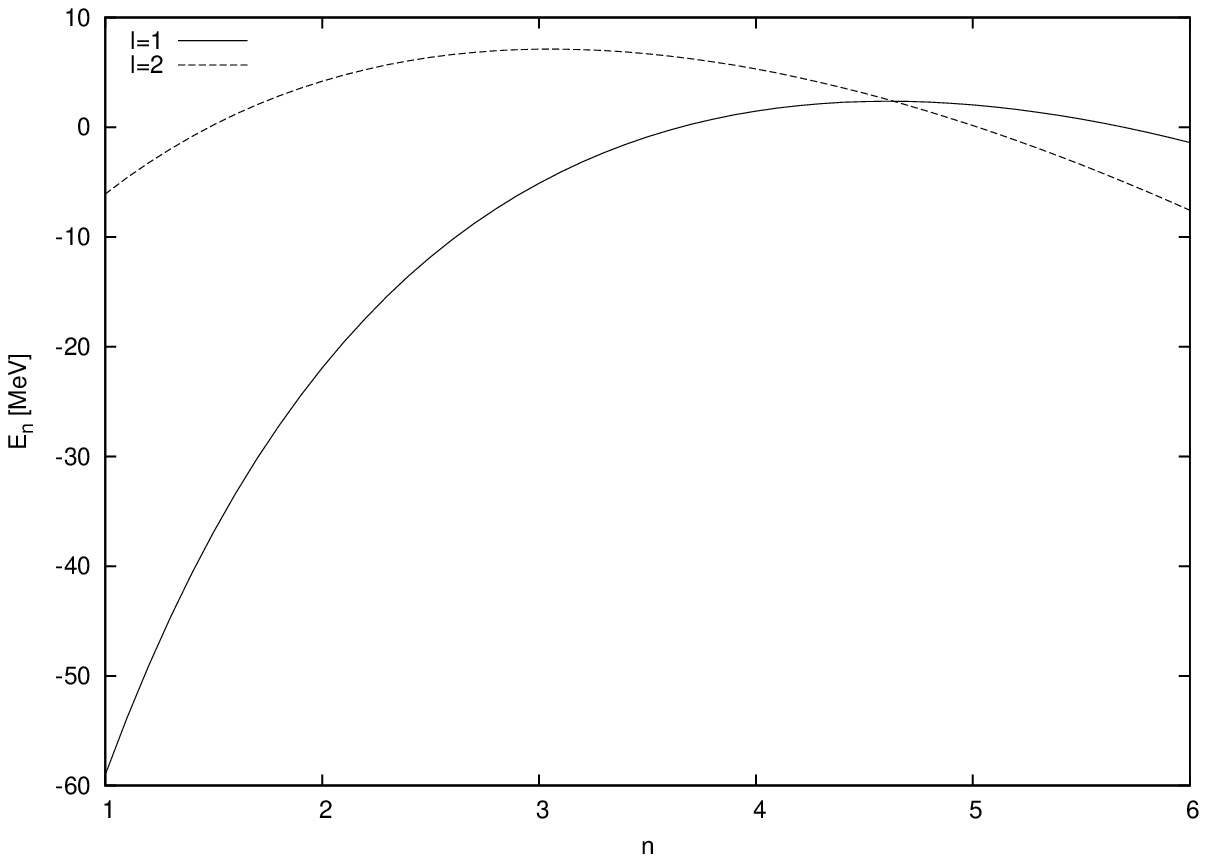}
\caption{The variation of the Schr\"{o}dinger energy levels with
respect to $n$ for $\ell=1, 2$.}
\end{figure}

\begin{figure}[htbp]
\centering
\includegraphics[height=2.5in, width=5in, angle=0]{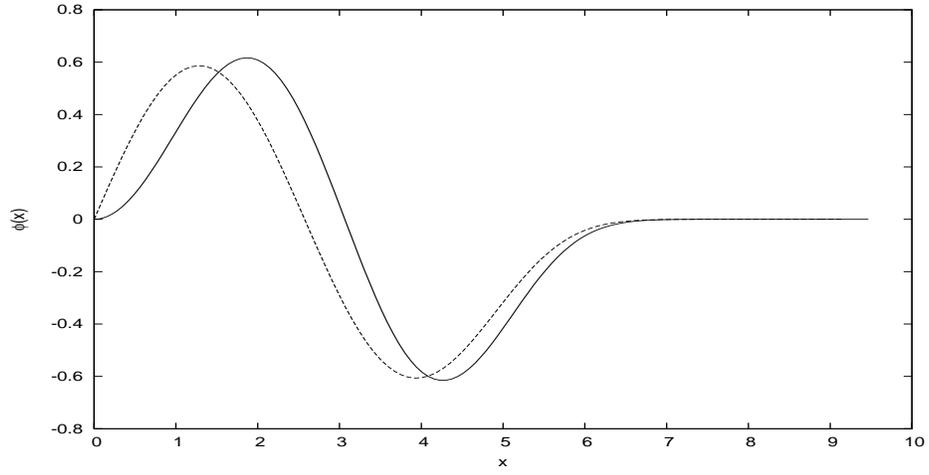}
\caption{The Schr\"{o}dinger eigenfunctions  without (full line),
and with using the approximation in Eq. (5) for Woods-Saxon
potential given in Eq. (57) for $n=1$\,, and $\ell=1$\,.}
\end{figure}

\begin{figure}[htbp]
\centering
\includegraphics[height=2.5in, width=5in, angle=0]{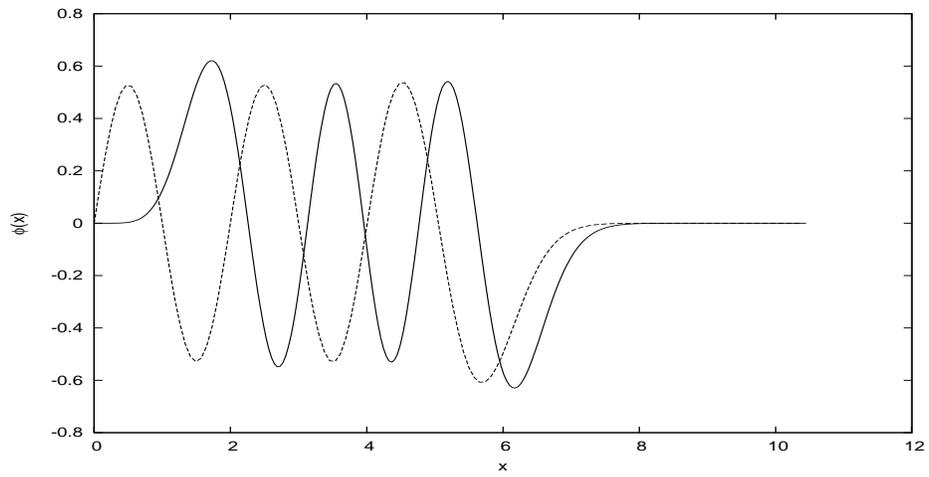}
\caption{The same as Fig. 7 but for $n=6$\,, and $\ell=5$\,.}
\end{figure}

\end{document}